\documentstyle[amsbsy,amstex,aps,epsf,multicol]{revtex}

\def\eqref#1{Eq.~(\ref{#1})}

\def\phi{\varphi}

\def\({\left(}
\def\){\right)}
\def\[{\left[}
\def\]{\right]}
\def\<{\left\langle}
\def\>{\right\rangle}
\def\<{\left\langle}
\def\>{\right\rangle}

\def\bea{\begin{eqnarray}}
\def\eea{\end{eqnarray}}

\def\pmb#1{\setbox0=\hbox{#1}%
  \kern-.025em\copy0\kern-\wd0
  \kern.05em\copy0\kern-\wd0
  \kern-.025em\raise.0433em\box0 }

\def\bsigma{\pmb{$\sigma$}}

\makeatletter

\title{L\'evy Model for Interstellar Scintillations}
\author{Stanislav Boldyrev${}^1$ and Carl R. Gwinn${}^2$\\
{\em ${}^1$Department of Astronomy and Astrophysics, University of Chicago, 
Chicago, IL 60637}, {\sf boldyrev@uchicago.edu} \\
{\em ${}^2$Department of Physics, University of California, 
Santa Barbara, CA 93106}, {\sf cgwinn@physics.ucsb.edu}}

\date{\today}

\begin{document}

\input psfig.sty


\maketitle

\begin{abstract}
Observations of radio signals from distant pulsars provide a valuable  
tool for investigation of interstellar turbulence. 
The time-shapes of the signals are the result of pulse broadening by the 
fluctuating electron density in the interstellar medium. 
While the scaling of the shapes with the signal frequency is well 
understood, the observed anomalous scaling with 
respect to the pulsar distance has remained a 
puzzle for more than 30 years.
We propose a new model for interstellar electron density fluctuations, 
which explains 
the observed scaling relations. We suggest that these fluctuations 
obey L\'evy statistics rather than Gaussian statistics, as assumed 
in previous treatments of interstellar scintillations. 


\end{abstract}

\begin{multicols}{2}

{\bf 1.} {\em Introduction.} Electron density fluctuations in the 
interstellar 
medium (ISM) cause
scintillations of the intensity of signals arriving from distant
pulsars. If the medium were completely transparent, the 
shape of the arriving signal would coincide with the shape of the 
signal emitted by the pulsar. However, the observed pulse is much 
broader, and this effect is attributed to the random refraction   
the waves experience while they travel through the 
medium~\cite{sutton,rickett1,rickett2,lee1,lee2,lee3}. To 
investigate pulse broadening  
one can assume 
that the pulsar {\em intrinsic} signal is narrow in time, 
$I_0(t)\propto \delta(t-t_0)$, where $I_0(t)$ is the signal intensity. 
The {\em observed} signal is broad and asymmetric,   
with a sharp rise and a slow decay;  
see Fig.~\ref{pulse}. 
Observations show that broadened shapes of the pulses are  
similar for different pulsars (after proper rescaling), 
suggesting that the density fluctuations statistics    
along different lines of sight are to some extent universal.

For estimates assume that the pulsar 
distance is $d\sim 10\,\mbox{kpc}$, the typical electron density is 
$n\sim 0.03\,\mbox{cm}^{-3}$, and the observational wave 
frequency is~$\nu \sim 500\,\mbox{MHz}$. 
Then the plasma electron frequency~$\omega_{pe}=(4\pi n e^2/m_e)^{1/2}$ 
is much smaller than $\nu$, and density fluctuations 
change the wave phase only slightly. To estimate the time delay 
one can use the approach of geometric optics, where the propagating 
ray is refracted (scattered) 
by small prisms of density 
inhomogeneities~\cite{williamson1,williamson2,blanford,boldyrev1,gwinn1}. 
At each scatter event 
occurring at a mean-free path $l$, 
the propagation angle changes by a small 
amount,~$\Delta \theta \sim \lambda^2 r_0 \Delta n$ [see below], 
where $\lambda$ 
is the wavelength, $r_0=e^2/m_e c^2$ is the classical 
radius of the electron, and $\Delta n$ is the density difference at  
characteristic separation~$l$.  Using the standard assumption that 
$\Delta \theta$ is random and Gaussian, one finds that the path direction 
deviates 
from a straight line by $\theta \sim \lambda^2 r_0 \Delta n_0 (d/l)^{1/2}$, 
where $\Delta n_0$ is the characteristic amplitude of density-difference 
fluctuations, and the path 
length deviates from the distance $d$ 
by $\Delta d \sim d \theta^2 \propto \lambda^4 d^2$.  
The broadening time can be estimated as  
$\tau_d\sim \Delta d/c$, which gives the 
standard scaling $\tau_d\propto \lambda^4 d^2$.

Observations show that the signal width, $\tau_d$, 
estimated at the half-amplitude level, scales with the wave length 
according to the obtained formula, $\tau_d \propto \lambda^4$, 
while the scaling with distance is close 
to $\tau_d \propto d^4$, contradicting the analytical prediction, 
as is seen in Fig.~2 \cite{hirano}.
This paradox was first discussed by Sutton~\cite{sutton}, and although the
theory of scintillations has been developed for more than 30 years, 
the contradiction has resisted analytical 
understanding~\cite{rickett1,rickett2}. 
{
\columnwidth=3.2in
\begin{figure} [tbp]
\centerline{\psfig{file=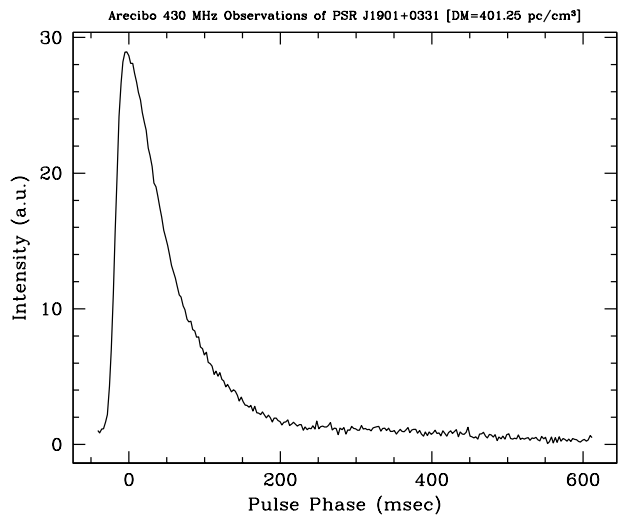,width=3.2in,angle=-0}}
\vskip3mm
\caption{Intensity of a typical observed pulsar signal averaged over many 
periods of pulsation. The shown time 
interval spans the pulsar period. The data were taken with the Arecibo 
telescope, at 430~MHz.  Courtesy of N. D. Ramesh Bhat~[7].} 
\label{pulse}
\end{figure}
}
In this paper we propose that the anomalous scaling 
with the distance is an evidence of {\em non-Gaussian} density 
fluctuations in the ISM. We suggest that the probability distribution 
of density gradients has a {\em power-law} decay, and its second 
moment is divergent. 
Such probability distributions are common in theories of turbulence, 
as is consistent with the argument that the density statistics are governed 
by turbulent motions 
in the ISM~\cite{lithwick,armstrong}. The sum of many angular 
deviations caused 
by such fluctuations 
does not have a Gaussian distribution; instead, the limiting distribution 
is of the L\'evy type, 
and the ray angle performs a L\'evy flight instead of a conventional 
random walk. We present a solvable model of scintillations that 
allows us to unify and extend 
to a non-Gaussian case the standard analytical approaches, 
see, e.g.,~\cite{williamson1,lee1}. We then apply this model to L\'evy density 
statistics,  compare it to the observational data, and demonstrate 
that the model naturally produces correct scalings of 
the signals.  We report main results here, the detailed 
discussion is presented in~\cite{boldyrev1}. 
{
\columnwidth=3.1in
\begin{figure} [tbp]
\centerline{\psfig{file=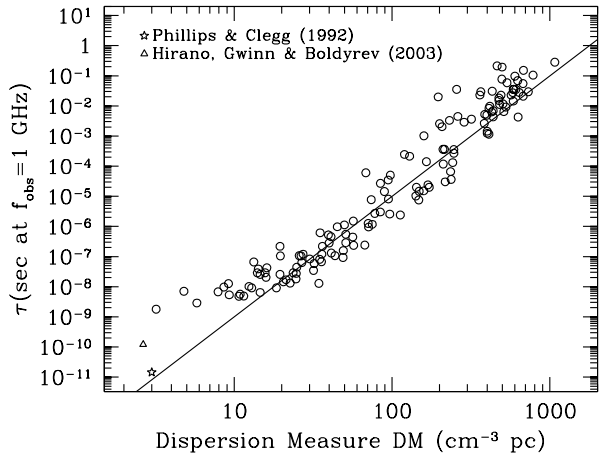,width=3.1in}}
\vskip3mm
\caption{Pulse temporal broadening as a function of 
the dispersion measure, $\mbox{DM}=\int_0^dn(z)\,\mbox{d}z$, which is a 
measure of the distance to the pulsar~[16].
Except as noted, data were taken from~[17].
The solid 
line has slope~4.}  
\label{elbow}
\end{figure}
}
{\bf 2.}{\em Wave equation in a random medium.}
The Fourier amplitude of electric (or magnetic) field, 
$E_{\omega}({\bf r})$, 
in the isotropic ISM with dielectric permittivity $\epsilon_{\omega}$ 
obeys the wave equation:
\begin{eqnarray}
\left[-\Delta -\frac{\omega^2}{c^2}
\epsilon_{\omega}({\bf r})\right] E_{\omega}({\bf r})=0,
\label{wave_equation}
\end{eqnarray}
where $\epsilon_{\omega}({\bf r})=1-\omega^2_{pe}({\bf r})/\omega^2$, 
and the electron plasma frequency $\omega_{pe}(\bf r)$ 
changes slowly on the wave 
scale~$\lambda$. Assuming that 
the wave propagates in the line-of-sight direction, $z$, we  
separate the quickly-changing 
phase of the wave from the slowly-changing amplitude,  
$E_\omega({\bf r})=\exp(iz\omega/c)\Phi_\omega (z,{\bf x})$, 
where ${\bf x}$ is a coordinate perpendicular to $z$.
Substituting this into the wave equation (\ref{wave_equation}), we 
derive the equation for the wave amplitude,
\begin{eqnarray}
\left[2i\frac{\omega}{c}\frac{\partial}{\partial z}+\Delta_{\perp}-
4\pi r_0 n({\bf x},z)\right]\Phi_\omega({\bf x}, z)=0,
\label{wave_equation2}
\end{eqnarray}
where $\Delta_{\perp}$ is a two-dimensional Laplacian in 
the ${\bf x}$~plane. Following \cite{lee1,lee2,lee3}, introduce the 
function  
$I({\bf r}_1, {\bf r}_2, t)=\Phi({\bf r}_1,t)\Phi^*({\bf r}_2, t)$, 
whose Fourier transform with respect to time is $I_\Omega({\bf r}_1, 
{\bf r}_2)=\sqrt{{2}/{\pi}}\int \mbox{d}\omega \,
\Phi_{\omega+\Omega/2}({\bf r}_1)\Phi^*_{\omega-\Omega/2}({\bf r}_2)$. 
For coinciding coordinates this function is the 
intensity of the radiation whose variation in time we seek. 
To find this function, we may first solve the equation 
for $V_{\omega, \Omega}({\bf r}_1, {\bf r}_2)\equiv
\Phi_{\omega+\Omega/2}({\bf r}_1)\Phi^*_{\omega-\Omega/2}({\bf r}_2) $, 
which can be derived from Eq.~(\ref{wave_equation2}). 
Assuming that $\Omega \ll \omega$, 
we obtain
\begin{eqnarray}
i\frac{\partial V}{\partial z}=
\frac{2k+\Delta k}{4k^2}
\frac{\partial^2 V}{\partial {\bf x_2}^2}
-\frac{2k-\Delta k}{4 k^2}\frac{\partial^2 V}{\partial {\bf x_1}^2}
+\frac{2\pi r_0}{k}\Delta n V,
\label{equation_V}
\end{eqnarray}
where we denoted $k=\omega/c$, $\Delta k=\Omega/c$, 
and $\Delta n=n({\bf x}_1, z)-n({\bf x}_2,z)$. 

Equation~(\ref{equation_V}) is 
hard to solve without further simplification  
since~$n({\bf x},z)$ is an unknown random function. The standard 
procedure is to assume that the density fluctuations are Gaussian 
with a specified correlator in~$x$ and only short-scale
correlations in~$z$, see~\cite{lee1}.  
Eq.~(\ref{equation_V}) can then be averaged 
over the Gaussian ensemble of density fluctuations and over 
different positions 
in ${\bf x}$-space. However, the resulting solution yields 
a scaling of $\tau_d \propto
\lambda^4 d^2$ that contradicts observations, as noted above.

We propose that the turbulent gas motions in the ISM give rise to
strongly intermittent and non-Gaussian density fluctuations. 
If 
the distribution function of $\Delta n$ has a power-law decay 
as $|\Delta n|\to \infty$ and has no second moment, then 
the sum of many independent ray angle deviations does not behave as 
a Gaussian variable (the central limit theorem does not hold). 
Instead, the limiting distribution, if it 
exists, is the L\'evy distribution. A random walk whose increments 
are L\'evy distributed is called a L\'evy flight.  
Such processes are common in various random systems, and 
often replace Brownian motion in turbulent systems~\cite{klafter}. 

The 
Fourier transform (the characteristic function) of a symmetric 
L\'evy distribution $P_{\beta}(\Delta n)$ has the simple form,
\begin{eqnarray} 
F(\mu)=\int\limits_{-\infty}^{\infty} \mbox{d}\Delta n\, P_{\beta}(\Delta n)
\exp(i\mu \Delta n)=\exp \left(-C|\mu|^\beta \right),
\label{levy}
\end{eqnarray}
where $0<\beta<2$, and $C$ is some positive constant. Eq.~(\ref{levy}) 
can be taken as the definition of a symmetric L\'evy distribution. The sum 
of $N$ L\'evy distributed variables scales 
as $\sum^N\Delta n\sim N^{1/\beta}$, which becomes diffusion 
in the Gaussian limit $\beta=2$. For $\beta < 2$, the 
probability distribution function has algebraic tails, 
$P_{\beta}(\Delta n)\sim |\Delta n|^{-1-\beta}$ for $|\Delta n|\to \infty$, 
and its second moment 
is divergent. We thus assume that the 
random density-gradient fluctuations are L\'evy distributed, 
and are short-scale correlated in~$z$. 
In the next section we first show how Eq.~(\ref{equation_V}) can 
be solved for a 
general case of non-Gaussian random density field. In Sec.~{\bf 4} we  
apply our method to L\'evy distribution.

{\bf 3.} {\em Batchelor approximation.}
Propagation as described by Eq.~(\ref{equation_V}) cannot
be simplified in general, when $\Delta n$ is not a
short-scale correlated Gaussian random variable.  However, analytical
investigation is possible in the important case of smooth turbulent
fluctuations.  This case is analogous to the Batchelor
limit, in the problem of turbulent random advection~\cite{frisch}. 
For that
approximation in this paper, we neglect all effects other than those
of density gradients: $n({\bf x}_1)-n({\bf x}_2)\simeq 
{\bsigma}(z)\cdot ({\bf x}_1-{\bf x}_2)$,  
where the density 
gradient~$\bsigma(z)$ is a random variable with correlation 
length~$l\ll d$ along $z$.  In this approximation, the variables 
separate in
Equation~(\ref{equation_V}) and it can be solved exactly. We leave analysis 
of more complicated cases for further communcation, and present here 
results for this simple case, which captures the essential physics.

As a further simplification, consider 
one-dimensional variables ${\bf x}_1$ and ${\bf x}_2$. Since the 
variables separate, we can 
look for the solution in the factorized 
form $V(x_1,x_2,z)=U_1(x_1,z)U_2(x_2,z)$. 
Then the equation for $U_1$ reads:
\begin{eqnarray}
i\frac{\partial U_1}{\partial z}=
-\frac{2k-\Delta k}{4 k^2}\frac{\partial^2 U_1}{\partial  x_1^2}
+\frac{2\pi r_0}{k} \sigma(z)x_1 U_1.
\label{simple_V}
\end{eqnarray}
The analogous equation for $U_2$ is obtained by changing $k\to -k$.
The solution of equation~(\ref{simple_V}) is sought in the form 
$U_1(x_1,z)=A(z)\exp\left[iB(z)x_1 + 
iC(z)x_1^2 \right]$, with the initial 
condition $U_1(z=0)= \delta(x_1)$, if the refracting medium 
extends all the way up to the pulsar. 
Substituting this ansatz into (\ref{simple_V}), we find 
\begin{eqnarray}
A(z)&=&\frac{A_0}{\sqrt{z}}\exp\left(\frac{-i(2k-\Delta k)}{4k^2}
\int\limits_0^z 
B^2(z') \mbox{d}z'\right),
\label{a1}\\
B(z)&=&\frac{-2\pi r_0}{kz}\int\limits_0^z\sigma(z')z'\mbox{d}z',\quad
C(z)=\frac{k^2}{(2k-\Delta k) z}\label{b1}.
\end{eqnarray}
Note that this solution describes the path of a single ray through a
sequence of density gradients~$\sigma(z)$.  Effects of multiple rays
can be found from superposition.  The intensity of received radiation 
can be calculated from  
the Fourier transform $I_{\omega}(z,t)=
\int_{-\infty}^{\infty}
 V_{\omega,\Omega}(x=0,z)\exp({-i\Omega t}) \mbox{d}\Omega/\sqrt{2\pi}.$
In this Fourier transform,
individual ray paths will yield contributions with phase proportional to
$\Omega=\Delta k c$, with coefficient equal to the travel time for that
path.  Cross-terms describing interference of paths yield
contributions that oscillate rapidly with frequency and average to 
zero, see e.g.~\cite{gwinn1}.  
The intensity, averaged over an ensemble of 
statistically independent rays, is then given by the average over 
individual travel times or 
equivalently over different realizations of~$\sigma(z)$.

This leads to 
\begin{eqnarray}
I_{\omega}(z,t)\propto 
\langle \delta\left(t-\frac{d}{c}-\frac{1}{2k^2c}\int\limits_0^z 
B^2(z') \mbox{d}z'\right)\rangle,
\label{shape1}
\end{eqnarray} 
where the angular brackets denote the statistical average.  
Note that $B(z)$ is proportional 
to the 
deflection angle $\theta$ of the ray. 
Formula (\ref{shape1})  gives 
the shape of the 
signal observed at the Earth; if the scattering medium were 
absent, the signal would be 
undistorted, $I(t)\propto \delta(t-d/c)$.

Lee \& Jokipii investigated Eq.~(\ref{equation_V}) for short-scale 
correlated 
Gaussian density fluctuations~\cite{lee1}. For averaging  
 over Gaussian $\sigma(z)$, 
our solution~(\ref{shape1}) reproduces 
 those obtained by 
Williamson~\cite{williamson1} with a phenomenological approach. 
Thus, Williamson's solutions are applicable 
under 
the assumption of smooth Gaussian density fluctuations, and when one keeps  
only the linear term in the expansion of~$\Delta n$. In the next 
section we apply our approach 
to the L\'evy distributed density fluctuations.

{\bf 4}. {\em Scintillations as L\'evy flights through the 
interstellar medium.} The averaging in formula~(\ref{shape1}) 
 can be performed for the L\'evy distributed short-scale 
correlated density 
gradients~$\sigma(z)$. To do this, we represent integrals 
is (\ref{b1}), and (\ref{shape1}) in the 
discretized forms, i.e., we   
assume that $d=nl$, $z'=ml$, and $z''=sl$, 
where $l$ is the correlation length 
of density fluctuations,  and change  
$\int_0^z f(z') \mbox{d}z' \to \sum_{m=1}^{n}f(l m)l $ for 
an arbitrary function~$f(z)$. The right hand side of 
(\ref{shape1}) 
is the probability distribution of the propagation time delay, 
$\tau=t-d/c$. 
For a continuous medium, this time delay is given by
\begin{eqnarray}
\tau=\frac{r_0^2l^3\lambda^4}{8\pi^2c}\sum\limits_{m=1}^{n}
\left[\frac{1}{m}\sum\limits_{s=1}^{m}s\,\sigma_s \right]^2. 
\label{tau_c}
\end{eqnarray}
The solution (\ref{shape1}) can now be 
calculated numerically as the probability distributions of this variable, 
under the assumption that $\sigma_s$ are distributed 
independently, identically, and according to the L\'evy 
law~(\ref{levy}). 
We however need to specify the parameter~$\beta$ in the L\'evy 
formula. We do this by comparing our model with observations. 
The observed scaling of pulse broadening is close to  
$\tau_d\propto\lambda^4 d^4$, while our model gives 
$\tau \propto \lambda^4 d^{(2+\beta)/\beta}$, as is seen from the 
scaling for sums of L\'evy-distributed variables, 
$\sum^m \sigma_s \sim m^{1/\beta}$, following Eq.~(\ref{levy}). 
Thus, we obtain,~$\beta \approx 2/3$. 

Note that standard Gaussian 
models of density distribution were not able to satisfy both observational 
scalings, $\lambda^4$ and $d^4$, simultaneously. Various studies of 
non-smooth density fluctuations within these models     
have not reproduced this scaling 
either~\cite{blanford,boldyrev1,goodman}. Our model reproduces the 
anomalous $d$-scaling naturally. Moreover, it predicts that the probability 
distribution function of electron 
density gradients in the interstellar turbulence   
decays as $P(\sigma)\sim |\sigma|^{-1-\beta} \sim |\sigma|^{-5/3}$. 
Power-law distributions $P(\sigma)$ with $\beta<2$ are indeed 
observed in numerical simulations of compressible turbulence~\cite{nordlund}, 
however, no one has yet 
derived them from first principles.
To date theories of scintillations have exploited only second-order 
correlators of the density fluctuations, 
while in our approach these correlators do not exist (or do not matter)    
and one must work with the whole probability distribution function. 

Although our goal was to explain the scalings of the signals, 
it is interesting to see to what extent we can predict their shapes. 
The delay time is proportional to the square 
of the typical deflection angle of the ray trajectory, 
$\tau\propto \theta^2$, where $\theta$ has the L\'evy 
distribution~$P_{\beta}(\theta)$. Therefore, the distribution 
of arrival times is $I(\tau)\propto P_{\beta}(\tau^{1/2})\tau^{-1/2}$, 
with the asymptotic form $I(\tau)\propto \tau^{-1-\beta/2}$ 
as $\tau\to \infty$. 
Fig.~(\ref{shapes})~shows the distribution  
of~$\tau$ from numerical calculation, with a power-law decay at 
long times, as expected. 
Because the observed shape of the scattered pulse is directly
related to the probability distribution of gradients in electron
density in the ISM, observational data offer the possibility of
characterizing interstellar turbulence~\cite{bhat}.
{
\columnwidth=3.1in
\begin{figure} [tbp]
\centerline{\psfig{file=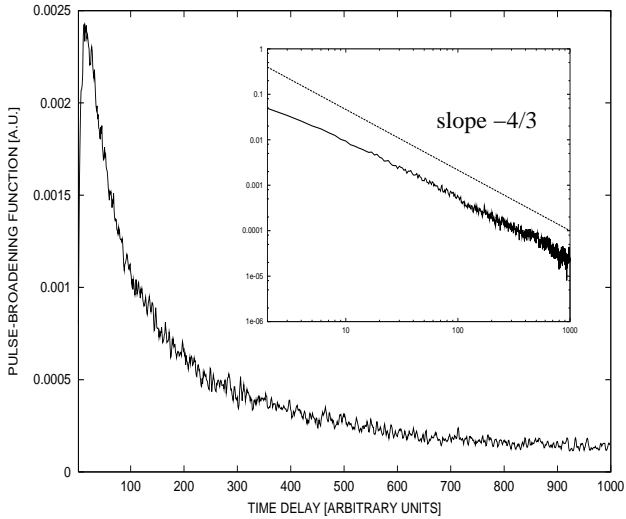,width=3.3in}}
\vskip3mm
\caption{Pulse-broadening functions for the model of linear 
density fluctuations obeying the L\'evy statistics with $\beta=2/3$. 
The distribution of the time delay (\ref{tau_c}) is 
found numerically using $10^6$ rays. The insert shows the large-time 
asymptotics 
of the curve in the log-log scale.} 
\label{shapes}
\end{figure}
}
The curve in Fig.~(\ref{shapes}) closely resembles the observed signals,
although the presented analytical shapes are the result
of averaging over an ensemble of non-interfering rays,
corresponding to an observational average over an infinite amount of time.
In practice, the averaging time is finite, and the long tail of the
distribution, dominated by rare events, may not have
converged.  
We also ignore instrumental 
response~\cite{bhat}.  
Moreover, 
a {\em non-analytic} density 
field is more natural
for a turbulent 
cascade~\cite{armstrong,lithwick,lambert}. 
Also, the small-scale 
density fluctuations that produce scintillations
should be collisionless, and 
elongated along the local magnetic field~\cite{lithwick}; the
scattering is likely to be highly anisotropic, and locally nearly
1-dimensional. 
We will consider these effects in 
future work. 
Interestingly, 
some scattered pulsars show 
power-law declines at long times, such as that in Fig.~\ref{pulse}. 
Scintillation of nearby pulsars also shows evidence for 
weak large-angle scattering~\cite{stinebring}.
Some interferometric studies suggest
a ``halo'' surrounding the source at large scattering angles
and excess scattering at small angles relative to a
Gaussian~\cite{gwinn3,desai},
as might be expected for a 
L\'evy distribution of scattering angles.
Intrinsic source structure, and the relatively short observatonal averages, 
may complicate this interpretation.

Finally, we wish to comment on the original explanation of the 
anomalous $d$-scaling by Sutton~\cite{sutton}. Sutton suggested  
that encounters with
much more-strongly-scattering HII regions become more probable on
longer lines of sight. This however requires a perhaps surprisingly 
close coordination 
of DM (over 1.7 orders of magnitude) with~$\tau_d$ 
(over 8 orders of magnitude). 
Sutton's proposal assumes essentially  
{\em non-stationary} statistics for the density 
distribution along $z$. Our proposal 
also invokes rare, large events, but in a 
statistically {\em stationary} way.

To summarize, we propose that the observed anomalously strong 
time-broadening 
of pulsar signals is evidence for non-Gaussian distribution of 
electron density gradients in the ISM. We argue that this 
distribution is of the L\'evy type, in accord with the turbulent 
origin of density fluctuations, and we 
present a simple model that explains the observational scalings of 
pulsar signals. 

The work of SB was supported by the ASCI Flash Center 
at the University of Chicago, under DoE subcontract B523820.

\end{multicols}

\end {document}